\documentclass[pre,twocolumn,showpacs,amsmath,amssymb]{revtex4}
\usepackage{graphicx}
\usepackage{epsfig}
\usepackage{dcolumn}
\usepackage{bm}

\begin{document}


\title{Effect of polydispersity and soft interactions on the nematic vs. smectic phase stability 
in platelet suspensions}

\author{Y. Mart\'{\i}nez-Rat\'on}
\affiliation{Grupo Interdisciplinar de Sistemas Complejos (GISC),
Departamento de Matem\'{a}ticas,Escuela Polit\'{e}cnica Superior,
Universidad Carlos III de Madrid, Avenida de la Universidad 30, E--28911, Legan\'{e}s, Madrid, Spain}

\author{E. Velasco}
\affiliation{Departamento de F\'{\i}sica Te\'orica de la Materia Condensada
and Instituto de Ciencia de Materiales Nicol\'as Cabrera,
Universidad Aut\'onoma de Madrid, E-28049 Madrid, Spain}

\date{\today}

\begin{abstract}
We discuss theoretically, using density-functional theory, 
the phase stability of nematic and smectic ordering in a suspension of
platelets of the same thickness but with a high polydispersity in diameter, and study the
influence of polydispersity on this stability. The platelets are assumed to
interact like hard objects, but additional soft attractive and repulsive interactions,
meant to represent the effect of depletion interactions due to the addition of non-absorbing polymer,
or of screened Coulomb interactions between charged platelets in an aqueous solvent, respectively, are
also considered. The aspect (diameter to thickness) ratio is taken to be very high, in order
to model solutions of mineral platelets recently explored experimentally. In this r\'egime
a high degree of orientational ordering occurs; therefore the model platelets can be taken
as completely parallel and are amenable to analysis via a fundamental-measure theory. 
Our focus is on the nematic vs. smectic phase interplay, since a high degree of polydispersity
in diameter suppresses the formation of the columnar phase.
When interactions are purely hard, the theory predicts a continuous nematic-to-smectic transition,
regardless of the degree of diameter polydispersity. However, polydispersity enhances the stability of the
smectic phase against the nematic phase. Predictions for the case where an additional soft 
interaction is added are obtained using mean-field perturbation theory. In the case of the 
one-component fluid, the transition remains continuous for repulsive forces, and the smectic phase 
becomes more stable as the range of the interaction is decreased. The opposite behaviour with respect 
to the range is observed for attractive forces, and in fact the transition becomes of first order
below a tricritical point. Also, for attractive interactions, nematic demixing appears, with
an associated critical point. When platelet polydispersity is introduced the tricritical 
temperature shifts to very high values.
\end{abstract}


\maketitle

\section{Introduction}

The issue of polydispersity is crucial to understand phase behaviour in experiments on colloidal 
suspensions of spherical and anisometric particles. The latter, of either
rod- or plate-like shape, are known to form liquid-crystalline phases \cite{1,2,3,4,5,6,7,10}
and their phase equilibria is largely affected by size dispersity \cite{11,11a,11b}. 
Colloidal particles can never be made truly identical, and even a small amount of polydispersity is 
unavoidable in real samples. 

Platelets with varying degrees of polydispersity, in particular, are being 
extensively studied in recent years.  In this material, polydispersity greatly facilitates 
gelation, and in fact the nematic phase is difficult to reach in 
suspensions of charge-stabilised platelike particles (the most common plate-like
particles) at particle concentrations below that at which a glassy state is formed \cite{gel}. 
However, it was possible to obtain a nematic phase in hard-like 
platelets, e.g. gibbsite particles with steric stabilisers \cite{6}. 
It has been demonstrated that the observation of equilibrium nematic phases in 
suspensions of charged colloidal platelets requires fine-tuning of the ionic 
strength of the solvent \cite{6,Lekker3}. Therefore, the concomitant effects of
polydispersity and soft interactions seem to be an important issue as regards
the stabilisation of the nematic phase.

Besides the nematic phase, liquid-crystalline phases with partial spatial order in colloidal suspensions of 
platelets are now being investigated \cite{Kooij}. The most common phase is the columnar phase, which admits
a surprisingly large (up to 25\%) degree of polydispersity in diameter. Suspensions of sterically-stabilised
gibbsite platelets with added non-absorbing polymer exhibit a rich phase diagram, with gravity-induced
three-phase coexistence regions involving isotropic, nematic and columnar phases, and enhanced fractionation 
effects \cite{Kooij2}. By contrast, observations of the layered
smectic phase are very rare in these systems \cite{Gabriel,Wang,Sun}, since the columnar phase
is very stable at high particle concentrations. 

In a recent paper \cite{Sun}, smectic ordering was
observed in colloidal suspensions of equally-thick charged platelets with high polydispersity
in diameter and in a solvent with low ionic strength. It is though that the extreme diameter 
polydispersity ($\agt 30\%$), possibly combined with the monodispersity in thickness, is
responsible for the instability of the columnar phase in favour of the smectic
phase. It would be interesting to theoretically analyse the effect of polydispersity on the formation 
of the spatially-ordered smectic phase in systems where the columnar phase is suppressed. 

On the other hand, additional soft interactions are known to affect the phase diagram.
These interactions can be screened Coulomb particle interactions due to the addition of
salt to an aqueous solution of charged particles, or attractive depletion forces \cite{AO}
between the colloidal 
particles arising from addition of non-absorbing polymer to a suspension of effectively hard platelets
\cite{vanD,Wensink,Luan,Petukhov}.
The possibility of theoretically predicting how both factors (polydipersity and soft
interactions)
influence the formation of phases with partial spatial order (smectic and columnar phases) would be 
highly desirable.

In the present paper we study the effects of polydispersity and soft interactions on the nematic vs.
smectic phase equilibrium. The work was inspired by the paper by Sun and coworkers \cite{Sun}, who
synthesised novel platelet particles from $\alpha$-ZrP minerals by
a special exfoliation procedure, which creates a perfectly monodisperse thickness distribution 
\cite{chinos,chinos2}
with high polydispersity in diameter. Low ionic-strength aqueous solutions of these charged 
particles form equilibrium nematic phases, followed by an incipient smectic ordering which 
equilibrates very slowly; no hint of columnar ordering was found \cite{Sun}. 

In the experiments by Sun et al. \cite{Sun} the ionic strength of the solution was not controlled.
However, addition of salt to the suspension will
modify the long-ranged repulsive interactions between platelets. Our second aim is to predict how the 
nematic vs. smectic phase equilibrium of the suspension will be modified when the range of
interactions is changed. These results could also apply to suspensions of
colloidal platelets with non-absorbing polymer, where effectively attractive depletion force between platelets usually 
result in demixing between phases with different concentrations of colloids and polymers \cite{Schmidt};
therefore, we also consider attractive soft interactions of varying range. As 
shown in Ref. \cite{Savenko}, addition of non-absorbing polymer to a one-component fluid of hard rods may give rise 
to isotropic liquid-liquid phase separation ending in a critical point (for large polymer coils), and to the presence 
of a broad coexistence region of isotropic-nematic, isotropic-smectic and isotropic-crystal phases. Binary mixtures
of colloidal disks and polymers exhibit not only isotropic-isotropic demixing, but also nematic-nematic demixing 
\cite{Bates}. However nonuniform phases were not taken into account in the latter study. As will be shown below,
soft interactions between highly oriented platelets can induce nematic-nematic demixing and a dramatic broadening of the 
nematic-smectic transition region for sufficiently long-ranged interactions.

The effect of polydispersity on the phase behavior of hard rods 
has been theoretically studied at the level of density functional theory (using the Onsager second virial theory) 
mainly for uniform liquid crystal phases (isotropic or nematic)
\cite{Clarke,Sollich1,Sollich2}. Also, a bimodal polydisperse distribution function in aspect ratio for hard board-like 
particles was used to study the effect of polydispersity on the phase stability of biaxial nematic vs. 
nematic-nematic demixing \cite{yuri0,yuri1}. The limit of zero polydispersity
gives rise to a perfect binary (bidisperse) mixture, a case studied by Varga and coworkers
\cite{Varga} using Onsager theory.  Harnau et al. \cite{Harnau} and 
Bier et al. \cite{Bier} have also studied perfect 
binary mixtures of platelets using density-functional theory \cite{Bier} in
the restricted-orientation approximation.
Finally, a simulation study was done by Bates and Frenkel on the influence
of polydispersity on nonuniform phases of rod-like particles:
it was shown \cite{11a} that the smectic phase of hard colloidal rods can be destabilized with respect to the columnar phase 
for high enough length polydispersity.

Density-functional studies of the relative stability of nonuniform (smectic, columnar and crystal) phases in 
polydisperse fluids made of anisotropic particles constitute a great challenge. The main difficulty is related to the 
numerical minimization of the functional as the density profile is a function of the spatial, angular and polydisperse 
variables. The present work is a first attempt to tackle the study of smectic phases in diameter-polydisperse platelets 
using density-functional tools and, in particular, fundamental-measure theory (FMT) of parallel hard cylinders \cite{Yuri-Cuesta}. 

We first give a brief overview of the theory in Sec. \ref{Theory}. In
Sec. \ref{hard} we examine the influence of polydispersity on the phase
behaviour of hard parallel platelets, while the case of soft platelets
is treated in Sec. \ref{soft}. We end with some conclusions in Sec. \ref{conclusions}.

\section{Model and theory}
\label{Theory}

The particle model consists of cylinders with parallel axes of revolution. This approximation 
is justified in the r\'egime of high packing fractions where the nematic-smectic (NS)
transition occurs (obviously the isotropic-nematic transition cannot be treated 
within this scheme, but this is not the aim of the present study).
Cylinders have the same thickness, $L_0$, in line with 
the particles obtained by Sun et al. \cite{Sun}, but are polydisperse in diameter. We assume the fluid 
to consist of a multicomponent mixture of species with different radii $R$.
In the thermodynamic limit the fluid contains infinitely many
species, and polydispersity can be characterised by means of a continuous size 
distribution. The particle radius  
will be characterised by the scaled variable $r=R/R_0$, 
with $R_0$ the mean radius. The number fraction of a given species with 
radius $r$ will be $p(r)$, the (continuous) radius distribution to be specified later.

\begin{figure}
\includegraphics[width=2.8in]{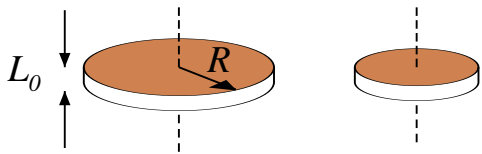}
\caption{\label{Fig0} Schematic of the parallel cylindrical particles used in this work.}
\end{figure}

\subsection{Density functional theory}

The statistical mechanics of a fluid of $N$ parallel cylindrical platelets in a volume $V$
will be investigated using the FMT density-functional approach for a multicomponent mixture of
parallel cylinders presented in \cite{Yuri-Cuesta}. Here we extend this approach to the polydisperse 
case. 

We choose the cylinder axes, the nematic director and the layer normal in the smectic phase to be 
parallel and along the $z$ direction. In the density-functional approach the central
quantity is the number density $\rho(z,r)$, which gives the local density at $z$ of the species 
with radius $r$ and satisfies
\begin{eqnarray}
\int_0^{\infty} dr \int_V d{\bf r}
\rho(z,r)=N.
\end{eqnarray}
Note that ${\bf r}=(x,y,z)$ refers to the space coordinate vector while $r$ is the 
scaled radius of particle (not the absolute value of ${\bf r}$). 
The excess part of the local free-energy density (assuming smectic symmetry) is
\begin{eqnarray}
&&\Phi(z)\equiv\beta f_{\hbox{\tiny ex}}(z)\nonumber\\
&&=-n_0(1-n_3)+\frac{2n_1^{\perp}n_2^{\perp}+n_1^{\parallel}
n_2^{\parallel}}{1-n_3}+\frac{n_2^{\parallel}\left(n_2^{\perp}\right)^2}
{(1-n_3)^2},
\label{Phi}
\end{eqnarray}
where $f_{\hbox{\tiny ex}}$ is the local excess free-energy density, 
$\beta=1/kT$, and $k$ is Bolztmann constant, with $T$ the temperature.
The functions $n_0(z)$, $n_1^{\parallel}(z)$, $n_1^{\perp}(z)$,
$n_2^{\parallel}(z)$, $n_2^{\perp}(z)$ and $n_3(z)$ 
[in (\ref{Phi}) their $z$ dependence has not been explicitly indicated]
are obtained from the number density $\rho(z,r)$ using the expressions:
\begin{eqnarray}
n_0(z)&=&\left[{\cal M}_0\ast \omega^{(0)}\right](z),\quad
n_1^{\parallel}(z)=\pi \left[{\cal M}_0\ast \omega^{(3)}\right](z),\nonumber\\\nonumber\\
n_1^{\perp}(z)&=&\left[{\cal M}_1\ast \omega^{(3)}\right](z),\quad
n_2^{\parallel}(z)=\left[{\cal M}_2\ast \omega^{(0)}\right](z),\nonumber\\\nonumber\\
n_2^{\perp}(z)&=&\pi \left[{\cal M}_1\ast \omega^{(3)}\right](z),\quad
n_3(z)=\pi \left[{\cal M}_2\ast \omega^{(3)}\right](z),\nonumber\\
\end{eqnarray}
where the symbol $\ast$ stands for convolution. The functions
\begin{eqnarray}
{\cal M}_{\alpha}(z)=\int_0^{\infty} dr \rho(z,r)R^{\alpha},
\end{eqnarray}
are generalized moments of the polydispersity distribution, while
\begin{eqnarray}
\omega^{(0)}(z)=\frac{1}{2}\delta\left(\frac{L_0}{2}-|z|\right),\quad
\omega^{(3)}(z)=\Theta\left(\frac{L_0}{2}-|z|\right),
\end{eqnarray}
with $\delta$ and $\Theta$ the Dirac delta and Heaviside functions, respectively. 
The excess free-energy functional is obtained by integration over
the system volume $V$, i.e. $\beta{\cal F}_{\rm{ex}}[\rho]=\int_V d{\bf r} \Phi(z)$.
Adding the ideal contribution $\beta{\cal F}_{\hbox{\tiny id}}[\rho]$, with
\begin{eqnarray}
\beta{\cal F}_{\hbox{\tiny id}}[\rho]=\int_0^{\infty}dr\int_Vd{\bf r}
\rho(z,r)\left\{\log{\left[\rho(z,r){\Lambda_r}\right]}-1\right\},
\end{eqnarray}
where $\Lambda_r$ is the thermal wavelength of the species with radius $r$,
we obtain the total free-energy functional as
${\cal F}[\rho]={\cal F}_{\hbox{\tiny id}}[\rho]+
{\cal F}_{\hbox{\tiny ex}}[\rho]$. The equilibrium state of the
system follows from functional minimisation of 
${\cal F}[\rho]$ with respect to the local number densities 
$\rho(z,r)$.

The radius distribution function $p(r)$ was chosen to be a Schultz
distribution,
\begin{eqnarray}
p(r)=\frac{(\nu+1)^{\nu+1}}{\Gamma(\nu+1)}r^{\nu}\exp\left[-(\nu+1) r\right],
\label{pr}
\end{eqnarray}
where $\nu$ is a free parameter that controls the width of the distribution.
Once $p(r)$ is chosen, the local density $\rho(z,r)$ must satisfy the
normalisation relation
\begin{eqnarray}
p(r)=\frac{\displaystyle\int_V d{\bf r}\rho(z,r)}{\displaystyle
\int_V d{\bf r}\int_0^{\infty} dr \rho(z,r)}=\frac{1}{N}
\int_V d{\bf r}\rho(z,r).
\label{distri}
\end{eqnarray}
Defining the moments as
\begin{eqnarray}
m_{\alpha}=\int_0^{\infty} dr p(r)r^{\alpha},
\end{eqnarray}
the distribution (\ref{distri}) fulfills
$m_0=m_1=1$. The polydispersity coefficient $\Delta$,
defined as usual as the width of the radius distribution, $\Delta=(m_2-m_1^2)^{1/2}/m_1$, 
is related to $\nu$ by $\Delta=(\nu+1)^{-1/2}$. In the smectic phase, local moments 
$m_{\alpha}(z)$ of the density distribution can also be defined:
\begin{eqnarray}
m_{\alpha}(z)=\frac{1}{\rho_0}\int_0^{\infty} dr r^{\alpha}\rho(z,r),
\end{eqnarray}
where $\rho_0=N/V$ is the mean density. Obviously
\begin{eqnarray}
\frac{1}{V}\int_V d{\bf r}m_{\alpha}(z)=
\frac{1}{d}\int_0^{d} dz m_{\alpha}(z)=m_{\alpha},
\end{eqnarray}
where $d$ is the smectic period.

\section{Hard platelets polydisperse in diameter}
\label{hard}

In the one-component case (platelets of the same diameter) the NS transition is located at a value of
packing fraction $\eta_s^{(0)}=0.314$ \cite{Capi}, where the packing fraction is defined as
$\eta=\rho_0 v$, with $v=\pi R_0^2L_0$ the mean particle volume and $\rho_0=N/V$ 
the mean density. The
transition is continuous. The question we would like to answer is whether the transition
is still continuous when polydispersity is introduced and how the packing 
fraction at the transition behaves as the sample polydispersity is changed.

\begin{figure}
\includegraphics[width=2.8in]{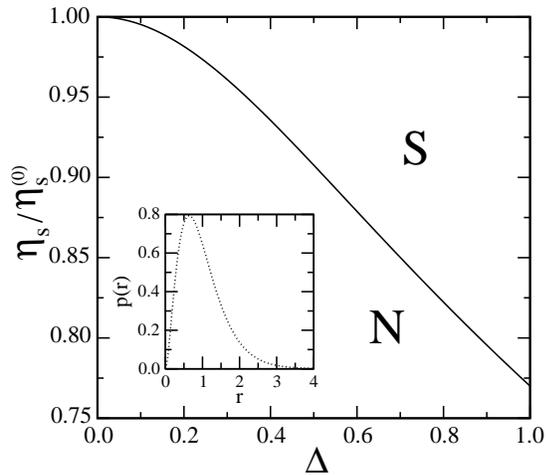}
\caption{\label{Fig1} Packing fraction $\eta_s$ at the NS spinodal, 
relative to that of the monodisperse fluid $\eta_s^{(0)}$, as a function of 
polydispersity parameter $\Delta$ for hard platelets. Labels indicate the nematic (N) and smectic (S) regions.
The inset is the radius distribution function at the spinodal for $\Delta=0.6$ }
\end{figure}

\subsection{Nematic-smectic bifurcation}

As a preliminary step prior to the full minimization of the functional,
we have calculated the spinodal (or bifurcation) line for the NS transition of the fluid. This can give us
an indication of the basic trends.
The calculation starts by perturbing the nematic with a small-amplitude
density wave, $\rho(z,r)=\rho_0 p(r) +\epsilon(r)\cos{kz}$, where 
$k=2\pi/d$ is a wavenumber. The fluid response
to the perturbation is obtained from the Fourier transform of the
direct correlation function (taken from the second functional
derivative of the excess free-energy functional):
\begin{eqnarray}
-\rho_0\hat{c}(k;r,r^{\prime})&=&c_0(k)(r+r^{\prime})^2\nonumber\\\nonumber\\&+&
c_1(k)rr^{\prime}(r^2+r^{\prime 2})+c_2(k)(rr^{\prime})^2,
\end{eqnarray}
where $r$, $r^{\prime}$ are the scaled radii of two platelets. Here we have defined the coefficients
\begin{eqnarray}
c_0(k)&=&y\left[2\chi(k)+ym_2\chi^2(k/2)\right],\nonumber\\\nonumber\\
c_1(k)&=&2y^2\left[2\chi(k)+(1+2ym_2)\chi^2(k/2)\right],\nonumber\\\nonumber\\
c_2(k)&=&y^2\left[2(1+2y)\chi(k)\right.\nonumber\\\nonumber\\&+&\left.(1+2y(2+m_2)+6m_2y^2)
\chi^2(k/2)\right],
\end{eqnarray}
where $\chi(k)=\sin{k^*}/k^*$ (with $k^*=kL_0$ the scaled 
wave number) and $y=\eta/(1-\eta m_2)$.
The instability 
of the fluid against the density wave is obtained from the eigenvalue problem
\begin{eqnarray}
\epsilon(r)=\rho_0 p(r)\int dr' \hat{c}(k,r,r')\epsilon(r'),
\label{ss}
\end{eqnarray}
from which the wavenumber $k_s$ and packing fraction $\eta_s$ at bifurcation
can be obtained. 

\begin{figure}
\includegraphics[width=2.4in]{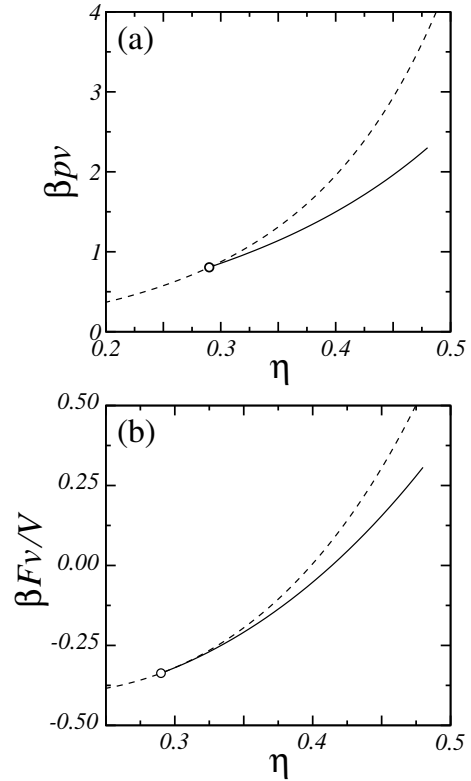}
\caption{\label{Fig2} (a) Scaled pressure $\beta Pv$, and (b) scaled 
free energy $\beta{\cal F}v/V$ as a function of packing fraction $\eta$ for the
nematic (dashed curve) and smectic (continuous curve) branches of hard polydisperse
platelets. The circle indicates
the location of the bifurcation point. The value of polydispersity is $\Delta=0.447$.}
\end{figure}

Fig. \ref{Fig1} shows the packing fraction of the NS spinodal, $\eta_s$,
relative to that of the monodisperse fluid, $\eta_s^{(0)}$,
as a function of polydispersity $\Delta$. The inset shows the radius 
distribution function for the particular value $\Delta=0.6$. The packing
fraction at bifurcation decreases as particles become more
polydisperse in radius, so the smectic phase becomes more stable
with respect to the nematic with increasing polydispersity. This trend
is opposite to that observed in fluids of hard rods, where
polydispersity in length postpones the onset of smectic stability due to 
incommensurability of the particle lengths with a smectic period. 

\begin{figure}
\includegraphics[width=2.6in]{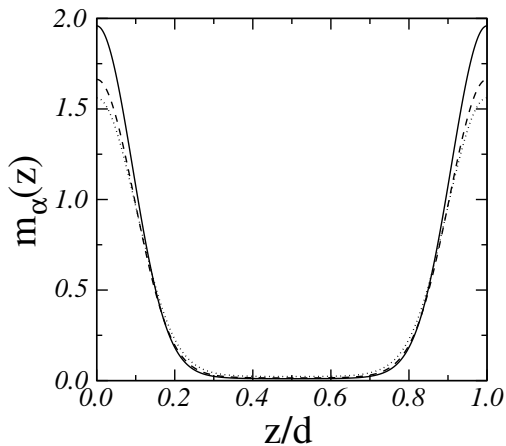}
\caption{\label{Fig3} Profiles of the moments $m_{\alpha}(z)$, with $\alpha=0,1,2$ (continuous,
dashed and dotted curves, respectively), along one
smectic period for hard polydisperse platelets in the case $\Delta=0.333$ and $\eta=0.455$. The smectic 
period is $d=1.212L_0$.}
\end{figure}

\subsection{Full minimisation and search for coexistence}
\label{full}

A full minimisation of the functional is required in order to gain more insight into 
the true nature (i.e. continuous vs. first-order)
of the NS transition and to obtain information 
on the equation of state, smectic period, and possible fractionation effects 
in the case $\Delta>0$ (i.e. the fact that the radius distribution function
can be different in the two coexisting phases).

We wish to calculate the density profile $\rho(z,r)$ of a stable smectic phase 
with a fixed number density $\rho_0$ and a prescribed radius distribution function $p(r)$. 
Note that the nematic phase will be unstable or metastable. In the minimisation of
the free-energy functional per unit volume, $\beta {\cal F}/V$, the density distribution is 
subject to the constraint $\rho_0p(r)=\left<\rho(z,r)\right>_d$ [see Eqn. (\ref{distri})],
with the shorthand notation$ \left<\cdots\right>_d\equiv d^{-1}\int_0^d(\cdots)$ for an
average over the smectic period. Direct functional minimisation leads to
\begin{eqnarray}
\rho(z,r)=\frac{\rho_0p(r)\exp{\left[c^{(1)}(z,r)\right]}}
{\left \langle \exp{\left[c^{(1)}(z,r)\right]}\right\rangle_d},
\label{eqnn}
\end{eqnarray}
with $c^{(1)}(z,r)$ the one-body correlation function. This function is a quadratic 
polynomial in $r$, with coefficients which are functionals of 
the moments $m_{\alpha}(z)=\rho_0^{-1}\int_0^{\infty}dr r^{\alpha}\rho(z,r)$. From
(\ref{eqnn}) we obtain a set of self-consistent non-linear integral equations in
$m_{\alpha}(z)$:
\begin{eqnarray}
m_{\alpha}(z)=\int_0^{\infty}dr \frac{r^{\alpha}p(r)\exp{\left[c^{(1)}(z,r)\right]}}
{\left \langle \exp{\left[c^{(1)}(z,r)\right]}\right\rangle_d},\hspace{0.2cm}\alpha=0,1,2.
\label{self-cons}
\end{eqnarray}
Expanding the local moments $m_{\alpha}(z)$ in a Fourier series containing the wavenumber
$k=2\pi/d$, Eqns. (\ref{self-cons}) can be written as self-consistent equations for
the corresponding expansion coefficients. The smectic period $d$ is obtained from the 
free energy by minimisation. For high packing fraction this process gives moments 
$m_{\alpha}(z)$ that are not constants: this corresponds to the smectic branch. Otherwise
the nematic branch is obtained.

Next we briefly outline how the conditions for NS coexistence were obtained. In equilibrium 
the chemical potentials of each species should be equal in both 
phases: $\mu^{(i)}(r)=\mu^{(0)}(r)$ ($i=\hbox{N,S}$ for nematic and smectic, respectively). 
$\mu^{(0)}(r)$ is calculated from the level-rule constraint
\begin{eqnarray}
\gamma\rho_{\hbox{\tiny N}}(r)+(1-\gamma)\rho_{\hbox{\tiny S}}(r)=\rho_0p(r),
\end{eqnarray}
which implies the conservation of the total number of particles; $\rho_0$ and 
$p(r)$ are the number density and radius distribution function of the 
parent phase. $0\leq\gamma\leq 1$ is the fraction 
of the total volume occupied by the N phase. Note that the number density  
of the S phase is $\rho_{\hbox{\tiny S}}(r)=\left<\rho_{\hbox{\tiny S}}(z,r)\right>_d$.  
Using the definition for chemical potential,
$\mu^{(i)}(r)=\delta {\cal F}/\delta \rho_i(z,r)|_{\rm{eq}}$, the equilibrium conditions 
together with the level rule provide a set of equations for 
the moments $m_{\alpha}^{(\hbox{\tiny N})}$ and $m_{\alpha}^{(\hbox{\tiny S})}(z)$:
\begin{eqnarray}
&&m_{\alpha}^{(\hbox{\tiny N})}=\int_0^{\infty} drr^{\alpha} p(r) 
\exp{\left[c^{(1)}_{\hbox{\tiny N}}(r)\right]}T(r), \label{mn}\\
&&m_{\alpha}^{(\hbox{\tiny S})}(z)=\int_0^{\infty} drr^{\alpha} p(r) 
\exp{\left[c^{(1)}_{\hbox{\tiny S}}(z,r)\right]}T(r),\label{ms}\\
&&T^{-1}(r)=\gamma\exp{\left[c^{(1)}_{\hbox{\tiny N}}(r)\right]}\nonumber\\&&
\hspace{2cm}+(1-\gamma)\left\langle\exp{\left[c^{(1)}_{\hbox{\tiny S}}(z,r)\right]}\right\rangle_d.
\label{tt}
\end{eqnarray}
These equations, together with the condition for mechanical equilibrium, 
i.e. equality of pressures of both phases $P^{(\hbox{\tiny S})}=P^{(\hbox{\tiny N})}$ [in our model
$\beta P=\left<\partial\Phi/\partial n_3\right>_d$], provide 
equilibrium moments of each phase and the number density $\rho_0$ of the parent uniform phase    
at coexistence.

A typical feature of polydisperse systems is that the composition of the two coexisting phases 
varies as the phase transition takes place. In our case this means the following.
Let us fix the value of $\Delta$ (the so-called dilution line). Then, as the total packing fraction $\eta$ of
the parent nematic phase is increased, first an infinitesimal amount of smectic 
material, called shadow smectic, will appear, coexisting with the cloud (parent) nematic phase. As
$\eta$ is further increased, the amount of smectic material will grow, and eventually the nematic will disappear.
The opposite process occurs by starting from a parent smectic phase and decreasing $\eta$; a smectic cloud point
will be reached when the first (shadow) nematic material can be observed in the sample.

The cloud-nematic and shadow-smectic coexistence curves (which are more relevant experimentally since one
usually starts from a parent nematic phase), can be calculated from 
Eqs. (\ref{ms}) and (\ref{tt}) by setting $\gamma=1$. This results in
\begin{eqnarray}
m_{\alpha}^{(\hbox{\tiny S})}(z)=\int_0^{\infty} dr r^{\alpha}p(r)e^{\Delta c^{(1)}(z,r)},
\hspace{0.4cm}\alpha=0,1,2,
\label{EL}
\end{eqnarray}
where $\Delta c^{(1)}(z,r)=c^{(1)}_{\hbox{\tiny S}}(z,r)-c^{(1)}_{\hbox{\tiny N}}(r)$ 
is the difference in the one-body correlation
functions of the two phases. In this case the nematic radius distribution function 
coincides with $p(r)$ (that of the parent phase). The smectic-cloud and nematic-shadow calculations  
can be implemented by setting $\gamma=0$ in Eqs. (\ref{mn}) and (\ref{tt}). In what follows, all coexistence results
correspond to the cloud-nematic and shadow-smectic curves for different polydispersities.


Our first result of the full minimisation is shown in Fig. \ref{Fig2}, where the scaled pressure 
$\beta Pv$ and the scaled free energy $\beta Fv/V$ are plotted as a function of packing fraction 
$\eta$ for the case $\Delta=0.447$. Calculations using many different initial guesses to solve
the self-consistent equations always lead to a single stable solution, which means that the nematic 
branch always bifurcates tangentially to a smectic branch. This implies that there is no coexistence,
and that the transition is always continuous. The scenario is the same for polydispersities larger 
than the one used in the calculations shown in the figure. 

It is instructive to examine the structure of the smectic phase by looking at
the profiles of the moments $m_{\alpha}(z)$. Fig. \ref{Fig3} plots the moments in one
smectic period for the case $\Delta=0.333$ and packing fraction $\eta=0.455$, i.e.
well inside the smectic region. As can be seen from the figure, the zeroth moment,
$m_0(z)$, reflects the structure of the total density, i.e. pronounced peaks located at the smectic
layers with a period of $d$. The first moment, $m_1(z)$, gives information about the mean 
particle radius at each location; at the layers the mean radius is $\simeq 1.65R_0$, and
it decreases away from the layers: larger particles are located preferentially at the
layers, whereas smaller particles tend to stay at the interstitials. 

\begin{figure}
\includegraphics[width=3.2in]{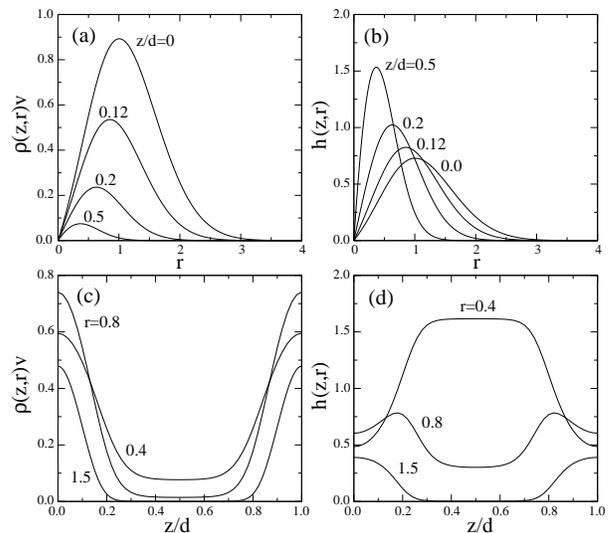}
\caption{Scaled density distribution $\rho(z,r)v$ and normalised
density profile $h(z,r)$ for a smectic phase of polydisperse hard platelets
with packing fraction $\eta=0.452$ and $\Delta=0.523$.
In (a) and (b), distributions are plotted as a function of the scaled radius $r$ for 
particular values of the $z$ coordinate, given as labels
(for reference, the smectic period is $d=1.211L_0$).
In (c) and (d), distributions are plotted as a function of the $z$ coordinate for
particular values of $r$.}
\label{Fig5}
\end{figure}

The latter effect can be studied in more detail by analysing the density profiles $\rho(z,r)$, 
along with the local radius distributions
\begin{eqnarray}
h(z,r)=\frac{\rho(z,r)}{\displaystyle\int_0^{\infty} dr\rho(z,r)},
\end{eqnarray}
as a function of $r$ for given $z$. This is shown in Figs. \ref{Fig5}(a) and (b)
for four different values of $z$. Complementary
plots are presented in Figs. \ref{Fig5}(c) and (d), where the same functions are given
as a function of $z$ for three particular values of $r$. The conclusion that can be drawn from
the figures is that there exists a microsegregation effect in the smectic phase. 
Specifically, particles with larger radii are preferentially located at the smectic layers, 
i.e. at $z=0$ or $d$, and possess a wider distribution function; by contrast, at the interstitial, $z=d/2$, the
distribution function $h(z,r)$ has a maximum at smaller values of $r$ and is much more narrow. 
These microsegregation effects become more pronounced as the degree of polydispersity is increased.

The increased stability of the smectic phase in fluids of hard polydisperse platelets,
shown in Fig. \ref{Fig1}, is due to the larger packing efficiency (lower excluded volume)
that platelets can achieve at the smectic layers when there is a continuous distribution of radii. 
As the distribution becomes
wider, and through a microsegregation mechanism, larger packing fractions can be achieved at the
quasi 2D-layers, which leads to a stabilisation of the layered arrangements with respect to the
uniformly distributed platelet configurations in the nematic phase.
In fluids of polydisperse rods, microsegregation also explains the destabilisation of the
smectic phase, since particles of different lengths cannot be arranged 
favourably into identical layers because of inefficient packing, and the minority species
is expelled to the interstitials.\\
%

\section{Soft platelets}
\label{soft}

Next we consider hard platelets that interact via an additional soft potential.
The soft potential will be treated by means of the usual mean-field approximation,
neglecting correlations. First we treat the case of a monodisperse fluid, leaving the general case
of a polydisperse fluid for the last section.

\subsection{Monodisperse fluid}

The soft potential should reflect the 
anisotropies of particle interactions. We have chosen a functional form
corresponding to a modified Yukawa potential where transverse and longitudinal
coordinates are scaled with the particle sizes $R$ and $L_0$, respectively:
\begin{widetext}
\begin{eqnarray}
V_{\rm{soft}}(r_{\perp},z)=\mp\epsilon
\left\{\begin{array}{ll}
\displaystyle\frac{e^{\displaystyle -\lambda\left[
\sqrt{(r_{\perp}/2R)^2+
(z/L_0)^2}-1\right]}}{\displaystyle\sqrt{(r_{\perp}/2R)^2+(z/L_0)^2}},&
\begin{array}{c}|z|>L_0,\\\hbox{or}\\
|z|<L_0\hspace{0.3cm}\hbox{and}\hspace{0.3cm}r_{\perp}>2R,\end{array}\\\\
0,&\hspace{1.2cm}\hbox{otherwise,}
\end{array}\right.\nonumber\\
\label{attr}
\end{eqnarray}
\end{widetext}
where $\epsilon>0$. Here $r_{\perp}$ is the interparticle relative distance 
(measured from the platelet centres) in the $xy$ plane, $z$ the relative 
distance along $z$, and $\lambda$ an inverse interaction range parameter.
Scaling the coordinates as in (\ref{attr}) is a natural choice, as it produces
ellipsoidal-like equipotential surfaces and discriminates between 
face-to-face and side-by-side configuration of two platelets, which should
have very different interparticle forces.

\begin{figure}
\includegraphics[width=2.8in]{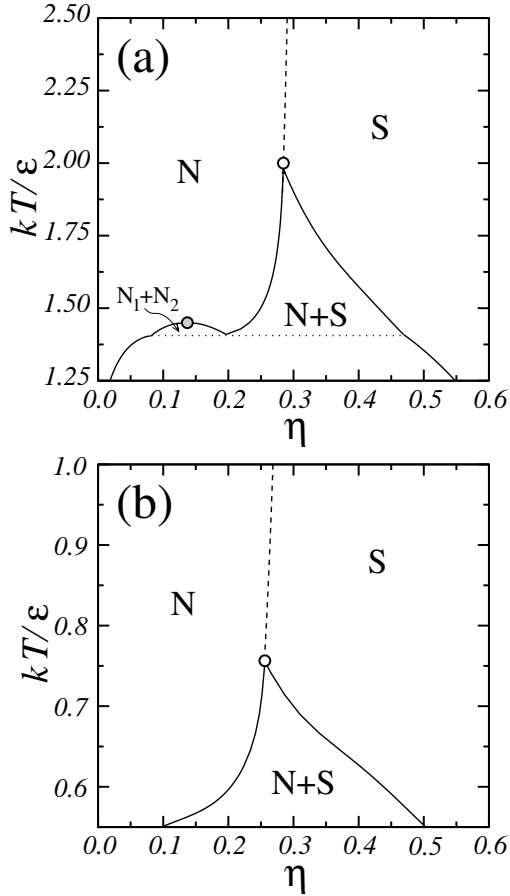}
\caption{Phase diagram of the monodisperse attractive Yukawa fluid. Continuous curves:
coexistence boundaries for the nematic-smectic or nematic-nematic transitions.
Dashed curves: nematic-smectic spinodal. Open circle: critical points. Shaded circle: tricritical point.
Dotted horizontal line: tieline of triple-point N$_1+$N$_2+$S coexistence. (a) $\lambda=1$. (b) $\lambda=2$.}
\label{cs1}
\end{figure}

The effect of the soft interaction on the free-energy functional is 
obtained from perturbation theory in the mean-field approximation:
\begin{eqnarray}
\frac{\beta {\cal F}_{\rm{mf}}}{V}=\frac{1}{2d}\int_0^ddz
\int_{-\infty}^{\infty}dz^{\prime}\rho(z)\rho(z^{\prime})u(|z^{\prime}-z|)
\label{uno}
\end{eqnarray}
with the effective potential
\begin{eqnarray}
u(z)=2\pi\int_{0}^{\infty}dr_{\perp} r_{\perp}\beta 
V_{\rm{soft}}\left(r_{\perp},z\right).
\label{ueff}
\end{eqnarray}
We use a Gaussian parametrization for the smectic density profile:
\begin{eqnarray}
\rho(z)=\rho d\left(\frac{\alpha}{\pi}\right)^{1/2}\sum_n\exp{\left[
-\alpha(z-nd)^2\right]},
\label{dos}
\end{eqnarray}
where $n$ is the set of integer values: $0, \pm 1, \pm 2,...$

In the present case of monodisperse particles, the procedure outlined in 
Sec. \ref{full} for obtaining the equilibrium structure of the fluid can be
made much simpler. Thus, for given values of scaled temperature $T^*=kT/\epsilon$
and interaction range parameter $\lambda$,
the calculation proceeds by fixing the value of the packing fraction $\eta$,
and then minimising the free-energy with respect to the Gaussian width
$\alpha$ and the smectic period $d$ to find the equilibrium density profile. 
Having obtained the free-energy branches as a function of $\eta$ for 
both nematic and smectic phases, the double tangent construction is used to
find the coexisting packing fractions. Changing $T$ and repeating 
the whole process, the phase diagram of the model in the 
$T-\eta$ plane is obtained. Considering the cases $-$ or $+$ 
in (\ref{attr}) one can cover both attractive and repulsive interactions.

\begin{figure}[h]
\includegraphics[width=2.4in]{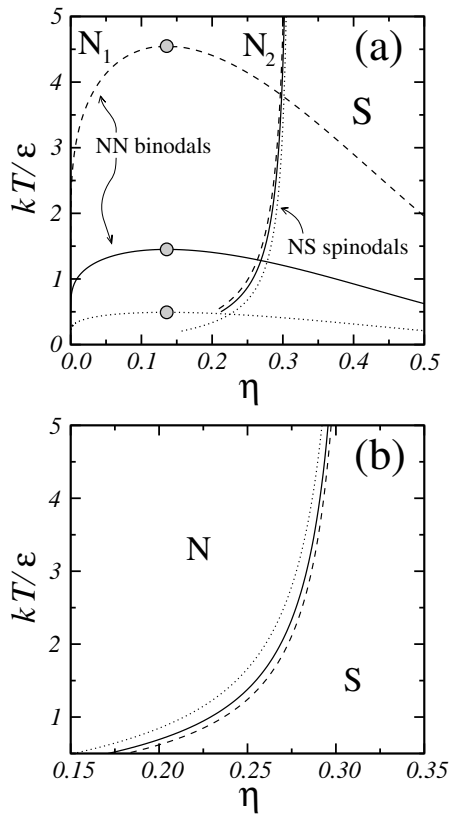}
\caption{Scaled temperature-packing fraction $T^*-\eta$ phase diagrams of the monodisperse model with (a) 
attractive, and (b) repulsive
soft interactions. Nematic (N) and smectic (S) phases are labelled by symbols. In the case
of demixing, two nematic phases, N$_1$ and N$_2$, coexist below a critical point,
indicated by circles. In both panels the inverse interaction length is
$\lambda=0.5$ (dashed curves), $1.0$ (continuous curves) and $2.0$
(dotted curves).}
\label{coex1}
\end{figure}

We now turn to the results. In the case of attractive forces, the NS transition
can be of first order or continuous, depending on the temperature. This is shown in Fig. 
\ref{cs1} for two values of the inverse range parameter $\lambda$. 
There is a tricritical point (indicated by an open circle in the figure), below which there is
NS coexistence. Above the tricritical temperature the transition is continuous.
This scenario is independent of the value of the interaction range. However, an interesting
feature of the phase behaviour is that, when the interactions are sufficiently long-ranged
(i.e. $\lambda$ sufficiently small), the nematic free-energy branch presents a region of 
instability, indicating the existence of nematic-nematic (N$_1+$N$_2$) demixing below a critical
temperature [with an associated critical point indicated by a shaded circle in 
Fig. \ref{cs1}(a)]. Obviously, as the inverse length decreases, nematic-nematic demixing 
becomes even more pronounced. This is shown in Fig. \ref{coex1}(a), where the N$_1+$N$_2$ demixing 
region in the $T-\eta$ plane is represented for different values of the inverse
lengths, $\lambda=0.5$, $1$ and $2$. The effect of the range of the interactions on the
smectic stability, which is relatively weak, can be appreciated from the change in the location 
of the NS spinodal as $\lambda$ is varied. As can be seen from Fig. \ref{coex1}(a), 
smectic ordering is slightly favoured as the range of the interactions is increased. 

The case of repulsive interactions is quite different. No nematic demixing transition exists 
in this case of repulsive interactions [Fig. \ref{coex1}(b)], and the 
NS transition is always continuous. Again the effect of varying the inverse length 
is weak, but different from the attractive case: for attractive forces, smectic ordering
is favoured as the interaction range is increased, but in the case of repulsive
forces, longer-ranged interactions disfavour the formation of a layered
structure with respect to a spatially uniform configuration. In any case, the
effect of the interaction range on the NS transition is larger for
repulsive forces.

\subsection{Polydisperse fluid}

When platelets are not equal in size several modifications in the
theory are necessary. The first is that density profiles depend on the polydispersity
variable $r$, and the soft potential $V_{\rm{soft}}(r_{\perp},z,r,r^{\prime})$ also 
contains a dependence via the radii of the two interacting platelets.
The mean-field free-energy contribution is now:
\begin{eqnarray}
\frac{{\beta {\cal F}_{\rm{mf}}}[\rho]}{V}&=&\frac{1}{2d}
\int_0^ddz\int_{-\infty}^{\infty}dz'\int_0^{\infty}dr\int_0^{\infty}dr'
\nonumber\\\nonumber\\&\times&\rho(z,r)\rho(z',r')u(|z-z'|,r,r^{\prime}),
\label{mf0}
\end{eqnarray}
where $u(z,r,r^{\prime})$ is defined as in (\ref{ueff}).
Next the soft interaction potential has to be specified. The soft potential will be defined as 
an extension of the one used previously, Eqn. (\ref{attr}), but for two platelets of different
radii $R=rR_0$, $R^{\prime}=r^{\prime}R_0$:
\begin{widetext}
\begin{equation}
V_{\rm{soft}}(r_{\perp},z,r,r^{\prime})=-\epsilon\left\{\begin{array}{ll}
\displaystyle\frac{\displaystyle e^{\displaystyle -\lambda\left\{\sqrt{
\left[r_{\perp}/(R+R')\right]^2+\left(z/L_0\right)^2}-1\right\}}}
{\displaystyle\sqrt{\left[r_{\perp}/(R+R')\right]^2+\left(z/L_0\right)^2}},&
\begin{array}{c}|z|>L_0\\\hbox{or}\\|z|<L_0\hspace{0.1cm}\hbox{and}
\hspace{0.1cm}r_{\perp}>R+R'\end{array}\\\\0,&\hbox{otherwise}
\end{array}\right.
\end{equation}
\end{widetext}
Again, since we are now considering a polydisperse fluid, a proper
calculation of the cloud nematic and shadow smectic curves has to be
implemented. The radius distribution function used in the following 
contains a Gaussian tail:
\begin{eqnarray}
&&p(r)=C_{\nu}r^{\nu}e^{-(s_{\nu}r)^2},
\end{eqnarray}
\begin{eqnarray}
C_{\nu}=\frac{2s_{\nu}^{\nu+1}}{\Gamma[(\nu+1)/2]},\quad 
s_{\nu}=\frac{\Gamma[(\nu+2)/2]}{\Gamma[(\nu+1)/2]}.
\end{eqnarray}
This choice is related to the better convergence reached 
in the numerical implementation of Eq. (\ref{EL}).

\begin{figure}
\includegraphics[width=2.8in]{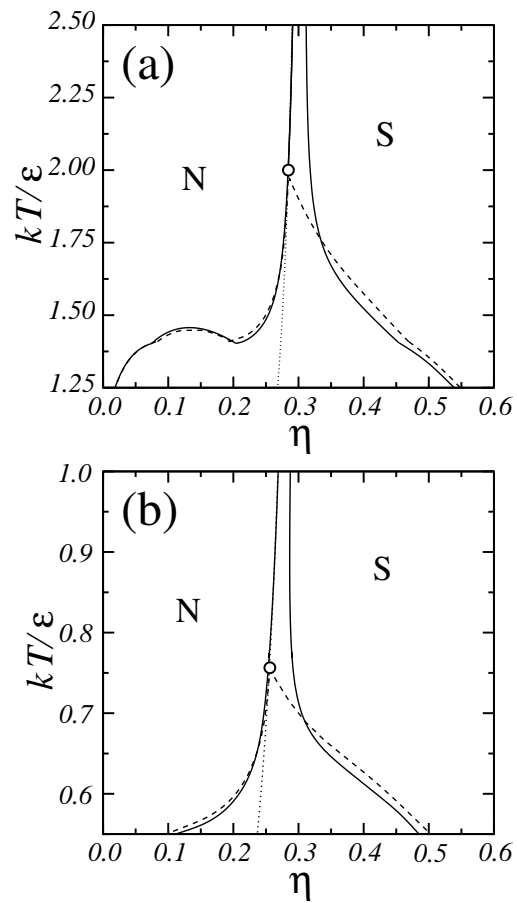}
\caption{Phase diagram of the polydisperse attractive Yukawa fluid. Continuous lines:
coexistence boundaries for the polydisperse fluid with $\Delta=0.294$. 
Dashed lines: monodisperse fluid. Dotted line: nematic-smectic spinodal.
(a) $\lambda=1$. (b) $\lambda=2$.}
\label{cs2}
\end{figure}

We have performed coexistence calculations and obtained complete phase
diagrams including nematic and smectic regions of stability, in this case
only for the attractive Yukawa-like potential (polydispersity is not expected 
to affect the phase behaviour significantly in the case of repulsive
forces). The phase diagrams are shown in Figs. \ref{cs2}(a) and (b), the first for 
an inverse length $\lambda=1$ and the second for $\lambda=2$. For the sake
of comparison, the results for the monodisperse fluid have been superimposed.
As far as N$_1+$N$_2$ coexistence is concerned, both the cloud and shadow curves of 
each nematic phase were calculated; however, their difference cannot be appreciated
at the scale of the figure, so only the cloud curve is plotted (this curve chosen
for consistency, as in the case of the NS coexistence only the cloud nematic-shadow 
smectic curve was computed). 
The most important change brought about by polydispersity is that the tricritical
point moves to very high values even for small values of $\Delta$, and therefore 
the NS phase transition becomes of first order in the reasonable 
range of temperatures. Again, for the longer-ranged potential, there exists an 
region of nematic demixing that is missing for shorter-ranged interactions,
a feature not much affected by polydispersity.

\begin{figure}
\includegraphics[width=3.5in]{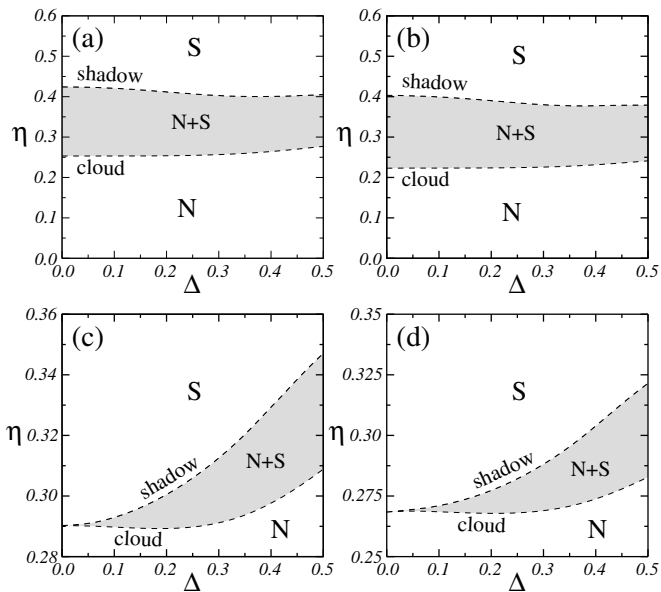}
\caption{Packing fraction vs. polydispersity for attractive Yukawa fluid
with: (a) and (c) $\lambda=1$; (b) and (d) $\lambda=2$.
Temperatures are: (a) $T^*=1.52$, (b) $0.63$ (both bellow corresponding
tricritical points), (c) $2.50$ and (d) $1.00$ (both above corresponding 
tricritical points).}
\label{effect}
\end{figure}

To show the effect of polydispersity on the NS transition in
attractive platelets and ascertain how the nature of the transition changes from the mono-
to the polydisperse case, we plot in Fig. \ref{effect} the coexistence 
packing fraction of the cloud-nematic and shadow-smectic curves as 
a function of polydispersity for $\lambda=1$ and $2$. The effect is weak
when the temperature is below the tricritical point of the monodisperse case,
since the density gap hardly changes with $\Delta$. By contrast, for temperatures 
above the tricritical-point temperature of the monodisperse fluid, and as the 
polydispersity is increased from $\Delta=0$, the density gap opens up quite
rapidly, meaning that the transition changes from first to second order even for a
low degree of polydispersity. This is shown in Figs. \ref{effect}(c) and (d). 

\begin{figure}
\includegraphics[width=3.2in]{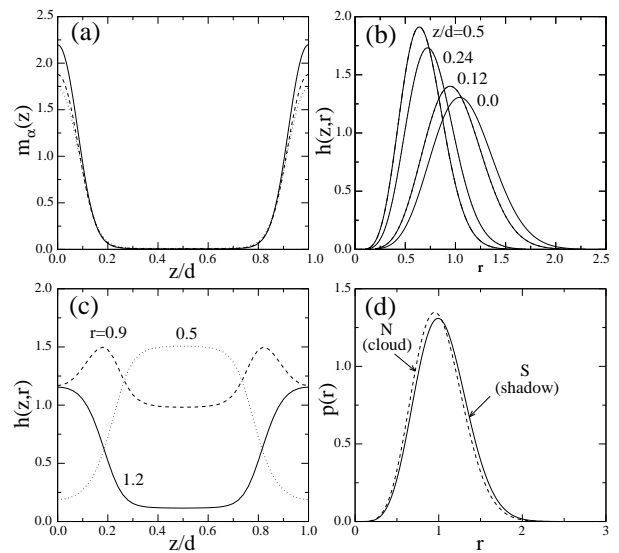}
\caption{Radius distribution functions in the (shadow) coexisting smectic phase for the
attractive polydisperse Yukawa fluid with $\lambda=2$ at scaled temperature
$T^*=0.6$ and degree of polydispersity $\Delta=0.294$. (a) Moments $m_{\alpha}(z)$
for $\alpha=0$ (continuous curve), $1$ (dashed curve) and $2$ (dotted curve).
(b) Size distribution function $h(z,r)$ as a function of $r$ for different locations
$z$ along one smectic period; from right to left $z/d=0$, $0.12$, $0.24$ and 
$0.5$ (the smectic spacing is $d=1.194L_0$). (c) Size distribution function $h(z,r)$ 
as a function of $z$ for different values of platelet radius; $r=1.2$ (continuous
curve), $0.9$ (dashed curve) and $0.5$ (dotted curve). (d) Radius distribution
functions $p(r)$ for the coexisting nematic (at the cloud curve) and smectic (at
the shadow curve).}
\label{micro}
\end{figure}

We have also studied the effect of microsegregation, i.e of the spatial redistribution
of particles with different sizes occurring along one period in the smectic phase
in the coexisting shadow-smectic phase. For this purpose, the moment distribution 
profiles $m_{\alpha}(z)$, defined previously, are useful. These functions are plotted in 
Fig. \ref{micro}(a), together with the distribution function $h(z,r)$ as a 
function of $r$ for fixed $z$ in (b), and as a function of $z$ for fixed 
$r$ in (c), all for $\Delta=0.294$ and $T^*=0.6$. The radius distribution functions
$p(r)$ in the N (cloud) and S (shadow) phases
are shown in Fig. \ref{micro}(d). The functions $m_{\alpha}(z)$ reflect the fact that
platelets are arranged in layers, with larger-sized platelets lying exactly at the layers
and with platelets being progressively smaller as one moves into the interstitial 
region. This microsegregation is more clearly appreciated in Fig. \ref{micro}(b),
where the size distribution function $h(z,r)$ is plotted for selected values of the
$z$ coordinate between the location of the layer $z=0$ and the intermediate distance
$z=d/2$. At the layers,
the maximum of the distribution occurs for larger values of the radius, whereas at the 
interstitial the maximum is located at a rather low radius. The width of the distribution 
decreases from the layer to the interstitial. The shift in average size as one moves
along a smectic period is again visible in Fig. \ref{micro}(c): of the three sizes
considered, larger particles ($r=1.2$) peak at the layers, and smaller ones at the 
interstitials, with intermediate sizes in intermediate positions.

The size distribution functions of nematic and smectic phases are quite similar,
with a slightly larger composition of the larger-sized platelets in the smectic
phase, but with almost identical widths. This trend is less and less apparent as
the polydispersity $\Delta$ is increased: nematic and smectic distributions 
become more and more similar, with mean radii that shift to lower values,
widths that become broader and tails that decay more slowly.

\begin{figure}
\includegraphics[width=2.0in]{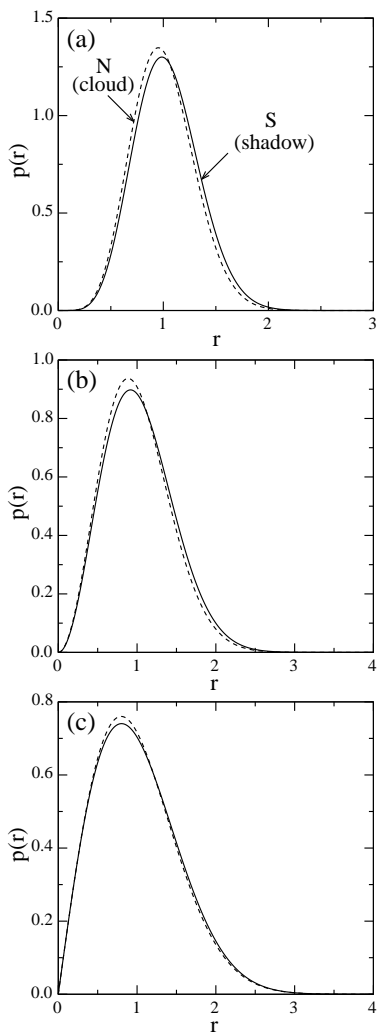}
\caption{Evolution of the radius distribution function $p(r)$ with increasing $\Delta$ for the
nematic in the cloud curve (dashed curves) and the smectic in the shadow curve (continuous
curves) at a scaled temperature $T^*=0.625$ and with an inverse length $\lambda=2$. 
(a) $\Delta=0.294$, (b) $0.422$ and (c) $0.523$.}
\label{diff}
\end{figure}

\section{Conclusions}
\label{conclusions}

In this work we have discussed the phase behaviour of suspensions made of platelets.
We were motivated by the recent finding of smectic ordering in an aqueous suspension of
$\alpha-$ZrP-based platelets of the same thickness but very polydisperse in diameter. The work has 
focused on the layered arrangement (smectic phase) of platelets arising from a suspension of 
positionally disordered, but perfectly orientationally ordered, particles (nematic phase). 
The high polydispersity in diameter prevents the formation of the columnar phase,
which has not been considered in the present work. The assumption of parallel, perfectly
oriented platelets should be valid when the aspect (diameter-to-thickness) ratio is
very high, which is the case in the $\alpha-$ZrP samples and in many other discotic
systems made from mineral materials. This assumption allows to use the powerful fundamental-measure
density-functional theory for hard platelets, which is expected to accurately describe ordered arrangement
of platelets, due to the emphasis of the theory on spatial correlations. Our 
first aim was to obtain a picture of how the degree of diameter polydispersity affects the nematic-smectic
phase transition. Polydispersity tends to stabilise the smectic phase, due to a microsegregation
effect which allows larger and similar particles to populate the layers and pack more efficiently.
The platelet volume fraction of the sample at the nematic-smectic transition can be reduced by
5\% for a degree of diameter polydispersity of 30\%, quite typical of the experimental system.
The transition is found to be always continuous, regardless of the degree of polydispersity.

Our second aim was to explore the consequences of soft interactions on the nematic-smectic transition.
Suspensions of platelets can be obtained using different techniques, and both attractive and
repulsive particles can be designed. We have contemplated both cases using a simple monotonic Yukawa-like
interaction potential function which can be positive or negative. The sign of the soft interactions 
profoundly affects the phase behaviour. For example, when platelets are identical (zero polydispersity),
soft repulsive forces always induce a continuous nematic-smectic transition, with the smectic phase
stabilising with respect to the nematic for a decreasing range of the force. By contrast, 
when the interactions are attractive, the transition becomes of the first order for temperatures
below a tricritical temperature. In addition, when the interaction range is sufficiently long,
there appears a nematic demixing phenomenon, whereby two nematic phases of different platelet concentration
appear in the suspension. Both nematic phases may coexist with a smectic phase at a triple-point
temperature. In this case the smectic phase is slightly destabilised as the range of the force
is decreased. Polydispersity changes this scenario to some extent. While nematic demixing is
not much affected (at least for polydispersities $\alt 30\%$), the tricritical temperature moves
to very high values, and the nematic-smectic transition
becomes of first order even for a small degree of polydispersity.

The application of the present results to real experimental samples may be done after
cautiously considering the type of interactions present in the platelets. Surface charges,
counterions, hydration effects, etc. may play a role in the total balance between attractive
and repulsive forces. Also, the interactions may have different sign with respect to distance.
Another aspect, not contemplated in the present work, is the role of the isotropic phase,
which is stable at lower packing fractions. The analysis of the isotropic phase requires
a different theoretical framework in order to account for orientational degrees of freedom. For 
example, the fundamental-measure density functional for freely-rotating hard platelets with vanishing thickness 
\cite{Schmidt1} (adequately
extended to the polydisperse case) could be a good 
candidate for a reference system in perturbation theories for soft platelets.  
A complete picture would be obtained by incorporating the columnar phase, which could in principle be done 
within the present theory, but at a higher computational cost. The computation of the full phase diagram,
and the study of how polydispersity in diameter and/or thickness, soft direct and solvent-induced interactions, 
etc., change the phase behaviour of platelet suspensions, is a big challenge from a theoretical point of
view. This work represents a step toward that goal.

\acknowledgements
This work has been partly financed by grants
NANOFLUID, MOSAICO and MODELICO from Comunidad Aut\'onoma de
Madrid (Spain), and Grants FIS2007-65869-C03-01, FIS2008-05865-C02-02, FIS2010-22047-C05-01 and 
FIS2010-22047-C05-04 from Ministerio de Educaci\'on y Ciencia (Spain).



\end{document}